# Resolving the Paradox of Changing El Niño-Monsoon Relation through Synchronization of Chaotic Oscillators

Devabrat Sharma[1,2,*], Sagar Sedani[1,2], Shruti Tandon[1,2], Gaurav Chopra[3], R. I. Sujith[1,2], and B. N. Goswami[4]

[1]Department of Aerospace Engineering, Indian Institute of Technology Madras, Chennai-600036, India
[2]Centre of Excellence for Studying Critical Transitions in Complex Systems, Indian Institute of Technology Madras, Chennai-600036, India
[3]Indian Institute of Technology Delhi, New Delhi-110016, India
[4]ST Radar Centre, Gauhati University, Guwahati-701014, India

*Devabrat Sharma
Email: devabratsharma9597@gmail.com

**Keywords:** Indian Summer Monsoon Rainfall, El Niño-Southern Oscillation, Long-Lead Predictability, Chaotic Synchronization.

**Abstract**

For over a century, the relationship between Indian summer monsoon rainfall and El Niño-Southern Oscillation has been the foundation of 'long-range' prediction of Indian monsoon. This relation is estimated from correlations between Indian summer monsoon rainfall and a Pacific sea surface temperature-based index of El Niño-Southern Oscillation. However, a prominent multi-decadal variability in the correlation raises doubts on the realism of El Niño-Monsoon relation and stability of Indian monsoon predictability. Previous studies discussed that Pacific-based El Niño-Southern Oscillation indices do not represent El Niño's global influence completely, making their correlation with Indian monsoon unreliable. To address this limitation, a Global El Niño-Southern Oscillation framework based on the depth of the 20°C isotherm is developed, integrating subsurface signal from all three tropical ocean basins and maximizing Indian monsoon teleconnections. Contrary to previous findings, the 20°C isotherm-based Global El Niño-Southern Oscillation exhibits a strong and stable correlation (>0.8) with Indian monsoon at an 18-month lead. Through a re-examination of the El Niño-Monsoon relationship with the superior Global El Niño-Southern Oscillation predictor, we show that the true relationship is independent of global warming and stationary in time. We discover that the stationarity in the El Niño-Monsoon relationship emerges as a natural consequence of chaotic synchronization between Indian summer monsoon rainfall and 20°C isotherm at an 18-month lead. Such synchronization between chaotic climate variables provides a new physical basis for climate predictability beyond the conventional deterministic limit set by chaos. Our findings provide the foundation and renewed confidence in long range climate prediction.

**Significance Statement**

Predictability of tropical monsoon is limited by virtue of the chaotic nature of climate subsystems. However, we show that this limitation is overcome when climate variables exhibit chaotic synchronization, thus radically improving the limit on predictability. Specifically, we examine the relationship between El Niño-Southern Oscillation (ENSO) and monsoon, that is traditionally examined through correlations between Indian summer monsoon rainfall (ISMR) and Pacific sea surface temperature indices. These correlations undergo temporal fluctuations and weaken at long lead times due to the chaotic coupled ocean-atmosphere system. Here, we discover that time-delayed synchronization occurs between chaotic climate signals associated with global sub-surface ocean variability and ISMR. We thus establish that the underlying ENSO-Monsoon relationship has remained robust over time, thus resolving current controversies.

## 1 Introduction

The relationship between Indian summer monsoon rainfall (ISMR) and El Niño-Southern Oscillation (ENSO) is well recognized as the basis for ISMR predictability. For over a century, the ENSO-Monsoon relationship has been quantified using the linear correlation between an ISMR index and a Pacific Sea Surface Temperature (SST) based ENSO index; for example, all-India June-September (JJAS) rainfall quantifying ISMR (Figure 1a), and similar seasonal mean SST averaged over equatorial Pacific (170°W-120°W, 5°S-5°N; Figure 1b) known as Niño 3.4 index representing ENSO (1–3). The ISMR-Niño3.4 correlation is significantly negative, fluctuating between -0.4 and -0.8 on decadal to multi-decadal timescales, with a mean of approximately -0.6. Notably, this relationship shows an overall declining trend in recent decades (1, 4, 5) (Figure 1c–f). While ENSO is considered the dominant predictor of ISMR (6–10), the reported post-1980s weakening of the ENSO-Monsoon relationship (4, 11) has raised concerns about ENSO's predictive role. Several studies have attributed this decline to natural versus forced variability (12–14), global warming (4, 11, 14), and changes in the mean states of the monsoon and ENSO (15).

However, the physical interpretation and quantification of variations of the ENSO-Monsoon relationship based on seasonal mean Pacific SST indices remains fraught with uncertainty, as both ISMR and Niño 3.4 contain substantial climate noise from high-frequency weather and sub-seasonal variability (16). Moreover, part of the epochal variability and apparent variability in ENSO-Monsoon relationship may arise from data gaps and sampling limitations (17). With improved SST analyses, the previously reported sharp decline in ENSO-Monsoon relationship (1, 4) appears weaker (Figure 1c-f), remaining within the multidecadal variability range, and even suggests modest strengthening post-2000s (18, 19). Furthermore, when ENSO-Monsoon relationship is evaluated using non-traditional predictor indices of ISMR such as the thermodynamic Indian Summer Monsoon index, the results show weak multidecadal variability and an increasing trend of relationship in recent decades (11). Similarly, the Pacific Warm Water Volume and ISMR correlation exhibits a weaker decreasing trend in recent years (20). These findings raise a pertinent question: Is the decadal-multidecadal variability of ENSO-Monsoon relationship real?

As ocean-atmosphere interactions over the Pacific are fundamental to ENSO genesis and evolution (21, 22), Niño 3.4 has naturally been used to characterize both ENSO variability and its teleconnection to regional climates such as the Indian monsoon. However, our recent study (23) identified two major issues in using Niño 3.4 as a predictor to quantify ENSO-Monsoon relationship. First, ISMR predictability is always underestimated due to 'climate noise' in Niño 3.4, making it unsuitable for long-lead predictions. Second, the influence of ENSO on ISMR at any lead time depends on the global tropical SST pattern at that lead. Thus, the regional Niño 3.4 predictor becomes insufficient to represent ENSO's impact on ISMR. Hence, the Niño 3.4-ISMR correlation represents only a partial teleconnection of ISMR with ENSO and could be erroneous at long leads. Recognizing these limitations and the importance of global tropical (0–360°E, 30°S–30°N) contributions, Sharma et al. (23), introduced the concept of 'Global-ENSO' (G-ENSO) as against 'Pacific ENSO' for estimating the correct strength of teleconnections.

Using the G-ENSO concept, we define a predictor (Dp) based on the depth of the 20°C isotherm (D20) that is significantly superior to predictors based on SST and unravels that the maximum potential predictability of ISMR is at 18-month lead (1). Therefore, it is imperative to examine whether the ENSO-Monsoon relationship contain the multi-decadal variability with the superior predictor at 18-month lead. Here, we investigate the role of uncertainty in observations on the ENSO-Monsoon relationship using 4 different rainfall and 4 different SST datasets. Using the Dp predictors, we unravel that the apparent multi-decadal variations of correlations between ISMR and SST based predictors is artificial, largely coming from a poor choice of predictors. We also provide a physical basis for the observed high correlation between ISMR and D20-based G-ENSO index at 18-month lead from nonlinear dynamics perspective. Further, we re-examine the ENSO-Monsoon relationship using G-ENSO and show that not only is ISMR more strongly coupled to G-ENSO than to Pacific ENSO, but also that ENSO-Monsoon relationship is not changing as perceived earlier. Remarkably, we find that the ENSO-Monsoon relationship was strong in the past and is likely to remain strong in the foreseeable future, even under a warming climate.

## 2 Results

### 2.1 A novel ISMR Predictor and the Changing ENSO-Monsoon relationship

Under the G-ENSO concept of teleconnections, we define two novel predictors of ISMR integrating the contributions from all three ocean basins based on monthly anomalies namely, the global tropical SST (Sp) and D20 (Dp). To construct Dp, first we obtain correlation maps between D20 and ISMR anomalies over a long time period (1875-2010) and retain correlations with >95% significance. Such correlation maps, constructed using D20 at different lead times, (Figure S1), represent the monthly evolutionary phases of ENSO in the Pacific Ocean and the associated patterns in the Atlantic and Indian Oceans. Next, using Hadamard (location-by-location) multiplication of the spatial pattern of D20 anomalies with the correlation map and then summing up across all locations, we obtain the time series of Dp. Similarly, we define Sp using SST anomalies.

There is evidence that simultaneously, the correlation of ISMR with both Indian Ocean Dipole Mode and Atlantic Niño tend to be opposite to that with Pacific ENSO index (24). The G-ENSO concept embraces simultaneous contributions from predictors such as ENSO in Pacific ocean, Indian Ocean Dipole Mode and Atlantic Niño within one framework. We thus refer to Sp and Dp, as G-ENSO based ISMR predictor indices and view Pacific ENSO as a dominant constituent of G-ENSO. To this end, we examined correlations of ISMR with monthly global tropical SST and D20 anomalies from 1- to 48-month leads (23).

The long-term correlation of ISMR with Sp and Dp as a function of lead month of forecasts are shown in Figure 1g. Important takeaways are, (i) Sp shows poorer potential skill at short leads than Dp, and further decreases with lead time of ISMR forecast. (ii) The potential skill of ISMR forecast shown by Dp increases with forecast lead time, reaching maximum long-term correlation >0.8 at 18-month lead, and (iii) the ISMR-Dp correlation at 18-month lead increases over time, reaching values greater than 0.9 in recent years (Figure 1h), possibly due to better sampling and analysis of D20 post Array for Real-time Geostrophic Oceanography (ARGO) era.

As the atmosphere responds to the global SST on timescales shorter than a season, a strong positive Dp-ISMR correlation at 18-month lead implies that Dp retains memory of the global SST anomaly pattern that would drive a strong monsoon after 18 months. We demonstrate this in Sharma, Das, Chakraborty et al. (25) using the regression of Dp and Sp, at zero and 18-month leads onto June-September mean SST and June-September meridional and vertical wind anomalies averaged over 70°E-90°E from 1000-200 hPa (Figure 2a-d). The meridional and vertical wind anomalies together represent the regional monsoon Hadley circulation. At zero lead, both Sp and Dp are associated with strong SST anomalies corresponding to a canonical La Niña condition in the equatorial Pacific, an Indian Ocean Dipole-like pattern in the Indian Ocean, and an Atlantic Niño signal (Figure 2a-b). These anomalies induce large-scale anomalous circulation response, with enhanced ascending motion and vigorous monsoon convection over the Indian subcontinent facilitated by increased moisture transport to land.

While Sp captures the essential tropical SST structure at zero lead, it loses predictive skill with increasing lead (Figure 2c). This decay reflects the influence of atmospheric stochastic variability on SST, resulting in a diminishing correlation between Sp and ISMR with lead time. In contrast, even at 18-month lead, Dp is associated with a strong La-Niña-like SST pattern and the strengthening of monsoon Hadley circulation (Figure 2d). The sub-surface D20 anomalies evolve slowly through a recharge–discharge process, providing persistent memory of ENSO precursors, largely unaffected by climate noise. This memory sustains the large-scale atmosphere-ocean circulation patterns (Figure 2d) necessary for the ISMR teleconnection, making Dp a robust predictor for skilful long-lead ISMR forecasts (23, 25, 26). We may not be able to use current dynamical coupled models to realize such 18-month lead predictability; however our deep learning model demonstrates the feasibility of achieving a correlation skill of 0.65 at 18-month lead over 44 years of independent ISMR forecasts (25).

### 2.2 Synchronization between ISMR and G-ENSO: Lag Synchronization between Coupled Chaotic Oscillators

Having established Dp as a superior G-ENSO predictor of ISMR (23, 25) (Figure 1g), here, we provide a physical basis for the 18-month lead ISMR predictability from nonlinear dynamics perspective. Seasonal prediction is evidently a slowly varying boundary forced problem (27), and not an initial value problem like weather. Yet, the spread in the simulation of seasonal mean prediction (28, 29) with perturbed initial conditions is an evidence of the influence of chaos even on the seasonal mean. In this backdrop, high potential predictability evident from ISMR-Dp correlation at 18-month lead may seem counterintuitive. However, studies in nonlinear dynamics have shown that when two chaotic oscillators are coupled, their dynamics can synchronize such that one oscillator follows a time-lagged, phase-lagged, or combined time- and phase-lagged version of the other, a regime known as chaotic synchronization (30).

Such chaotic synchronization can provide a basis for ISMR predictability far beyond what was thought possible based on the error growth associated with the system. To demonstrate this, we defined generalized G-ENSO indices that are completely independent of the ISMR-derived correlation map. A G-ENSO index may be defined as the dominant mode of monthly mean global tropical ocean variability. It thus provides a more comprehensive measure of the large-scale ocean-atmosphere interactions that modulate tropical circulation and, consequently, ISMR. By definition, the first principal component of the leading empirical orthogonal function of a tropical field represents the dominant mode of coherent tropical variability of that field. Figure 3a-b indicates that the first principal component of June-September SST (hereafter S-Nino) correlates strongly with Sp at zero lead month (~0.99; Figure 3a), and first principal component of June-September D20 (hereafter D-Nino) correlates strongly with Dp at zero lead month (~0.80; Figure 3b). Therefore, S-Nino and D-Nino can be interpreted as generalized counterparts of the ISMR-specific G-ENSO indices or generalized G-ENSO indices.

As scalar time series such as D-Nino, S-Nino, and ISMR arises from coupled ocean-atmosphere system with inherent memory and feedback processes, each of these variables represent only a one-dimensional projection of a potentially multidimensional dynamical system (31). Therefore, when the underlying system dynamics evolve in a nonlinear multidimensional state space, phase space reconstruction (see Materials and Methods; Figure S2) provides a framework to recover the underlying dynamics of multidimensional systems from a single time series (32, 33). Further, the intrinsic dynamical properties of the D-Nino and ISMR trajectories in the phase space, exhibiting complex and chaotic behavior, can be quantified using joint recurrence analysis (see Materials and Methods) (34). Recurrence methods, widely used in physiology (35–37), engineering (38–42), and climate science (43–46), quantify the geometry of trajectories in reconstructed phase space and are particularly well suited for relatively short time series (31) (~100 data points). By characterizing how trajectories revisit previously occupied regions of the phase space, recurrence plot provides insight into the degree of synchronization in the evolution of nonlinear dynamical system (see Materials and Methods). A recurrence plot is represented as a binary matrix of ones and zeros, where each one (black dot) indicates that the system at time $i$ is closer to its state at time $j$ than a prescribed threshold distance ($\varepsilon$), determined based on the recurrence rate, in the phase space.

The element-wise product of the recurrence plots of two dynamical systems forms the joint recurrence plot, which captures the instances when multiple systems simultaneously return to their respective past states (Figure 4a-b; see Materials and Methods). The joint recurrence plot between ISMR and D-Nino/S-Nino, with 20% recurrence rate, provides key insights into the joint evolution of the systems (Figure 4a-b). The strength of the dynamical coupling between the systems under consideration can be quantified by the density of simultaneous recurrences (1s or black dots; Figure 4a-b) in the phase spaces, referred to as joint recurrence rate. Figure 4c shows that ISMR and D-Nino exhibit statistically significant joint recurrence rate at lead times of 4-6 months and 16-18 months, indicating stronger dynamical coupling between them. For ISMR and S-Nino, the joint recurrence rate also indicates strong dynamical coupling at multiple lead times (Figure 4e).

However, one of the most prominent features in joint recurrence plots is the presence of diagonal line structures (47) that indicate intervals during which ISMR and D-Nino/S-Nino evolve coherently along comparable dynamical pathways, indicating phase synchronization (48) between the systems under consideration. Hence, longer diagonal segments in the joint recurrence plot at a given lead time indicate stronger phase synchronization between ISMR and D-Nino/S-Nino at that lead time. In contrast, scattered recurrence points and shorter diagonal lines without coherent structures, are an indication of the systems' joint evolution being dominated by randomness or high-dimensional noise even with strong coupling (48). The maximum length of diagonal in ISMR-S-Nino joint recurrence plot remains comparatively weak and largely statistically insignificant relative to the ISMR-D-Nino relationship (Figure 4c and 4f). Hence, the long-lead predictability of ISMR at 16-18 months lead arise from strong dynamical coupling and phase synchronization between the inherently chaotic ISMR and D20 oscillatory system. Consistent with the findings of Sharma et al. (23), the joint recurrence analysis also indicates that the interaction between ISMR and SST-based indices is less deterministic and more strongly influenced by stochastic variability, limiting its effectiveness for estimating ISMR predictability.

Although phase synchronization provides the physical basis for the high long-lead predictability of ISMR, it is insufficient to quantify the potential predictability limit of ISMR. ISMR exhibits significant multidecadal variability (Figure 5a). For high predictability, predictors at a given lead must exhibit similar multidecadal variability and remain in phase with ISMR. Therefore, in addition to phase synchronization, a lagged relationship must also exist between the amplitudes of variability of the predictor and ISMR to quantify predictability. This form of chaotic synchronization is known as lag synchronization (30, 49), in which the states of two dynamical systems become nearly identical but are separated by a finite time delay, such that $Y(t) \approx X(t-\tau)$. Thus, lag synchronization between two chaotic systems is characterized by two key signatures: (i) a delayed correspondence in their amplitudes of variability, and (ii) a consistent phase relationship between their oscillatory components.

Interestingly, the predictor discovery algorithm developed by Sharma et al. (23) can effectively identify the hidden phase synchronization between ISMR and a predictor without requiring phase-space reconstruction. Consequently, the amplitude variability of the predictors derived from the algorithm can be used to detect lag synchronization with ISMR and estimate its predictability. The weak correlation between ISMR and Sp at zero lead reflects the absence of significant multidecadal variability in Sp (Figures 5a and 5b). Although Dp at zero lead exhibits multidecadal variability (Figures 5a and 5c), it is not in phase with ISMR. In contrast, at 5- and 18-month leads, Dp shows stronger phase locking with ISMR (Figures 5d–f). In particular, the scatter between ISMR and Dp(-18) follows an approximately linear relationship, consistent with $Y(t) \approx X(t-\tau)$ (Figure 5f). As a result, the amplitudes of ISMR and Dp(-18) align systematically at the lagged time, producing the observed strong correlation. Furthermore, the high phase-locking value (see Materials and Methods) (50) of 0.75 between ISMR and Dp(-18) satisfies the key signatures of lag synchronization, thereby providing the basis for estimating the true potential predictability of ISMR at long leads.

### 2.3 ENSO-Monsoon Relationship and ISMR predictability in a Warming Climate

Next, we examine if the ENSO-Monsoon relationship is indeed changing as indicated in some studies for the historical period (1850-2020) using Pacific ENSO index (4, 11, 18, 19). We re-examine the evolution of the ENSO-Monsoon relationship using Sp and Dp at three leads. A 21-year moving correlation of ISMR-Sp (Figures 6a-c) reveals a non-stationary behaviour, with alternating epochs of stronger and weaker correlations over decadal-multidecadal timescales. Uncertainty increases with lead time due to climate-noise-driven error growth in Sp. While zero-lag correlations are relatively consistent across rainfall datasets, the spread among data sets widens at 5- and 18-month leads, and the long-term ISMR-Sp correlation declines from ~0.6 at zero lead to ~0.2 at 18-month.

In contrast, ISMR-Dp correlations (Figures 6d-f) behave differently. At zero month lead (Figure 6d), correlations fluctuate with multidecadal variability and are non-stationary similar to Sp (Figure 6a). With increasing lead time, however, the amplitude of epochal variations decreases (Figure 6e) and at 18-month lead the ENSO-Monsoon relationship stabilizes, showing minimal epochal variability and consistently strong positive correlations across datasets (Figure 6f). The long-term correlation increases from 0.6 at zero lead to 0.85 at 18-month lead. These results are consistent with the joint recurrence-based synchronization analysis. The strong synchronization between ISMR and D-Nino at 18-month lead, shown by the large maximum diagonal length (Figure 4d), indicates coherent and persistent joint evolution of ISMR and D20 over extended time period, thereby leading to the reduced epochal variability in the ISMR-Dp(-18) relationship. In contrast, shorter diagonal structures in ISMR-S-Nino relationship at 5- and 18-month lead (Figure 4f), leads to multidecadal variability in the ISMR-Sp relationship

We hypothesise that the long-term decreasing trend in the ENSO-Monsoon relationship with Sp (Figure 6a) is due to climate noise which is absent in the ENSO-Monsoon relationship with Dp (Figure 6d). This indicates that the multidecadal variability in the ENSO-Monsoon relationship with both Sp and Dp at lead of zero-month is due to artificial variability introduced in both Sp and Dp from sampling issues and data gaps. The question arises, how does Dp act as a filter at long-leads, reducing the influence of potential artificial variability from analysis deficiencies in D20 that may lead to variability in the ENSO-Monsoon relationship? A satisfactory answer would require additional investigation and ocean analysis sensitivity experiments, presently outside the scope of this study. However, our intuitive understanding is the following.

D20 represents the thermocline depth and is governed by the equatorial dynamics in deep tropics (5°S-5°N), while the off-equatorial thermocline depth between 5°-30° latitude (both hemisphere) could be influenced by extra-tropical dynamics such as Pacific Decadal Oscillation or Atlantic Multidecadal Variability. Poor constraints (from data gaps, sampling, and resultant analysis) on the equatorial Kelvin and Rossby waves driving the air-sea interaction on interannual time scales resulting in poor constraints on the slow thermocline wave may lead to climate noise in Dp, thus explaining the epochal variations of ENSO-Monsoon relationship with Dp at zero lead (Figure 6d). The long-term ISMR-D20 correlation map (Figure S1) shows that off-equatorial D20 correlates strongly with ISMR at all leads. Also notable is that the deep tropical D20 between 5°S and 5°N is dominated by interannual variations (Figure S3a). However, when off equatorial D20 is included, the spectrum of D20 averaged between 20°S-20°N, shows that the interannual variability of D20 is modulated by a significant multi-decadal variability with period ~50 years (Figure S3b). At short leads, within the equatorial adjustment period, the equatorial D20 contributes more to Dp resulting in epochal variations of ENSO-Monsoon relationship. However, at leads of 18-month and beyond, the off-equatorial D20 contributes more to Dp. At such long leads, beyond the ocean adjustment period, the climate noise may be minimal and multi-decadal modulation from off-equatorial D20 provides the necessary memory needed for Dp to lag synchronize with climatic variables such as ISMR.

How do the Pacific Decadal Oscillation or the Atlantic Multidecadal Variability introduce decadal-multi-decadal variability on D20 in tropics (10°N-30°N)? Studies indicate that extra-tropical thermal and wind forcing can influence the tropical thermocline through subduction (51). The low-level basin-scale anticyclonic circulation over the North Pacific, associated with ISMR multidecadal variability (Figure 5a of Rajesh & Goswami (52)) or with the Atlantic Multidecadal Variability (Figure 12 of Rajesh & Goswami (52)), likely forces a stationary anomaly in the subtropical Pacific gyre, establishing a multidecadal quasi-stationary D20 anomaly between 10°N and 30°N. The Atlantic Multidecadal Variability is also known to contribute to seasonal ISMR predictability, and the atmospheric bridge for this teleconnection has been well-documented (52–56). Detecting the Atlantic Multidecadal Variability signature in tropical SST has been challenging due to the dominant ENSO signal and stochastic SST noise. We argue that stationary atmospheric Rossby waves (52, 53) can generate quasi-stationary off-equatorial oceanic Rossby waves, introducing multidecadal variability in the off-equatorial thermocline, which is captured by Dp at 18-month lead.

### 2.4 G-ENSO-Monsoon Relationship in Projections

Conventional wisdom suggests that as extreme weather events intensify and become more variable under climate change, the seasonal ISMR predictability would steadily decline in coming decades (57). On the contrary, we find that ISMR predictability remains largely unaffected during the historical period, despite $CO_2$ rising by over 50% and methane by more than threefold relative to preindustrial levels (58). However, whether this will hold under further greenhouse gas forcing remains to be established. To explore this, we examined ISMR-Dp relationship from Coupled Model Intercomparison Project Phase 6 (CMIP6) simulations under representative concentration pathways 2.6 and 4.5 scenarios. Representative concentration pathways of 2.6/4.5 refers to simulation runs leading to an extra 2.6/4.5 Watts of additional heat trapped per square meter of Earth's surface compared to the pre-industrial era due to greenhouse gases.

Figures S4a-b shows Dp-ISMR correlation across models at multiple leads, and Figure 6g-h shows 21-year moving Dp-ISMR correlation at 18-month lead under both scenarios. While substantial inter-model variability is evident in the long-term correlation strength, the correlation remains high and is devoid of epochal variations up to 2 years in advance (Figure S4a-b) as in the historical period (Figure 6f). Weak correlation in some models is likely due to systematic biases in ENSO representation, the annual cycle of ISMR, their coupling, and errors in simulating ISMR multidecadal variability (59). Despite this, at least two models maintain robust, statistically

significant correlations of 0.6–0.8 at long leads under representative concentration pathway 2.6, and three models show correlations of 0.6-0.9 under representative concentration pathway 4.5, closely matching historical observations (Figure S4a-b). Unlike SST-based indices (Niño 3.4), the Dp-ISMR correlation, particularly at 18-month lead, exhibits little multidecadal variability during the historical period. Figures 6g-h show that at least one high resolution CMIP6 model (Table S1) reproduces this feature in future projections, maintaining strong correlations at multiple leads even under elevated greenhouse gas forcing. Overall, our analysis demonstrates that the Dp-ISMR teleconnection remains robust in both historical period and selected CMIP6 projections. While model biases affect the precise magnitude and timing of correlations, the persistence of correlations for long leads in most models (Figure S4a-b) underscores the importance of sub-surface oceanic predictor such as D20 for ISMR prediction. These findings supports our previous conclusion from historical data that ISMR predictability is expected to remain robust under future climate scenarios.

## 3 Discussion

Our results provide new insight into the physical basis of long-lead predictability of ISMR. Although individual components of the climate system exhibit chaotic behavior, their high-dimension nonlinear dynamics can become synchronized at certain relative lead times, providing a basis for long-lead predictability. Synchronization of coupled chaotic oscillators is well established in dynamical systems theory (30, 60–66) and has been explored to characterize phase relationship among climate variables (45, 67, 68). However, the existence of such chaotic synchronization between climate variables is unprecedented knowledge. The joint recurrence approach presented in this work, offers a powerful tool for uncovering hidden predictability and can be extended to investigate long-lead predictability in other climate systems. The recurrence quantification analysis performed using a stricter recurrence rate of 10% (Figure S5) yields qualitatively similar results, indicating that the conclusions are robust to the choice of recurrence threshold.

Given that the effective embedding dimension of the phase space of these coupled climate oscillators (ISMR, D-Nino, and S-Nino) is of the order of ~6-7, a sufficiently long time series is required to reliably characterize their recurrence structure and mutual interactions. Here, a data set length of ~130 years provides the requirement for robust phase-space reconstruction and recurrence-based analysis. Importantly, such a record encompasses both interannual and multidecadal variability present in ISMR, SST, and D20. This is crucial, because meaningful phase synchronization and hence predictability requires coherence across these multiple timescales. In particular, for any predictor to exhibit high skill at long leads, it must remain dynamically synchronized with ISMR not only at interannual scales but also across multidecadal variability. Therefore, the use of ~130 data points ensures that the analysis adequately captures the full spectrum of variability governing the coupled system and supports a physically consistent interpretation of long-lead predictability.

Since Gilbert Walker's early work (69), the ENSO-ISMR relationship has been estimated using equatorial Pacific SST indices, treating it as ideal predictor. Regional SST-based predictors such as the Indian Ocean Dipole, Pacific Decadal Oscillation, and Atlantic Niño, often assumed independent of ENSO, have also been used as ISMR predictors at various leads. Our study shows that these assumptions are premature, as regional SST indices contain substantial climate noise, sampling error, and represent only a partial measure of the connection between global tropical oceans and the Indian monsoon. Hence, the previously reported epochal variations in the ENSO-Monsoon relationship is largely due to the over reliance on regional SST predictors.

Through impact on tropical thermocline, Pacific Decadal Oscillation and Atlantic Multidecadal Variability impart a synchronous multidecadal variability with approximately 65-year periodicity (Figure S3) on Dp at 18-month lead, explaining its high correlation with ISMR. Our findings unravel that ISMR predictability need not be restricted by the limit on deterministic predictability if a predictor has high correlation with ISMR at a longer lead. This is due to nonlinear lag synchronization between ISMR and its G-ENSO predictor, a fundamental property of the ENSO-Monsoon relationship. As a consequence, the relationship shows negligible sensitivity to historical climate change or to projections under increased greenhouse gas forcing. These findings alleviate the perceived potential danger from ISMR predictability decreasing significantly in the near future.

The G-ENSO framework developed in Sharma et al. (23) transforms how we understand and quantify teleconnections in Earth's climate. Since G-ENSO represents the dominant global mode of tropical variability, we expect all tropical monsoon systems, Australian, East Asian, African, and American, to be phase-locked to the seasonal cycle and governed by the same global mode. Predictors analogous to Dp can therefore be developed for each monsoon system to assess their long-lead predictability. In the climate system, interactions between weakly chaotic oscillators are ubiquitous, indicating that potential chaotic lag synchronization can also be ubiquitous. So far, extracting a reliable orderly signature of high predictability amidst the background chaos has been a formidable challenge hitherto. In a first, our construction of Dp provides a recipe to identify such nonlinear lag synchronization in the climate system from observations.

## 4 Materials and Methods

### 4.1 Data

The Indian Summer Monsoon Rainfall (ISMR) is defined as the total accumulated rainfall over the Indian landmass during June-September (JJAS). To illustrate the uncertainty in the ISMR indices, we show the time series of four ISMR indices from four rainfall products: (i) the Indian Institute of Tropical Meteorology (IITM) dataset, based on 306 fixed stations and available for 1871-2016 (70); (ii) the European Centre for Medium-Range Weather Forecasts Reanalysis version 5 (ERA5), gridded at 0.25° × 0.25° resolution and available from 1940 to the present (71); (iii) the India Meteorological Department (IMD) gridded dataset at 1° × 1° resolution, available from 1901 to 2023 (72) which is based on a fixed network of 1384 rain gauge stations, and (iv) the IMD Climate Monitoring Portal (referred to as IMD2), also covering 1901–2023 (73).

The Niño 3.4 index, a measure of estimating the strength of the Pacific ENSO is measured using SST anomalies averaged over 170°W-120°W, 5°S-5°N. We show the time series of four Niño 3.4 indices from four SST products: (i) Centennial In-situ Observation-Based Estimates of SSTs (COBE), gridded at 1° × 1° and available from 1850 to the present (74); (ii) the Hadley Centre Global Ice and Sea Surface Temperature dataset (HadISST), gridded at 1° × 1° and available from 1870 onward (75); (iii) the Extended Reconstructed Sea Surface Temperature dataset (ERSST), gridded at 2° × 2° and spanning from 1854 to the present (76); and (iv) the Kaplan Extended SST version 2 dataset, gridded at 5° × 5° and available for 1856–2022 (77). The larger uncertainty in Niño 3.4 indices in the earlier historical period (Figure 1b) seems to be consistent with 'sampling error' during the data sparse period, while the larger uncertainty in the ISMR indices in recent decades (Figure 1a) may be related to larger 'climate noise' due to larger spatio-temporal variability of rainfall from global warming in different data sets. The variability in climatological mean and the standard deviation of ISMR and Niño 3.4 index across multiple datasets is shown in Table S2.

The 20°C isotherm depth (D20) over the tropics is obtained from the Simple Ocean Data Assimilation (SODA) version 2.2.4, gridded at 0.5° × 0.5° and available for the period 1871–2010 (78). Monthly meridional and vertical wind velocities at multiple pressure levels, on a 0.25° x 0.25° global grid from 1940 to 2010, were obtained from ERA5 for regression analysis. D20 and ISMR simulations under representative concentration pathways of 2.6 and 4.5 are obtained from Coupled Model Intercomparison Project Phase 6 for the period 2015-2100 (Table S1). For all datasets, anomalies are computed by removing the climatological mean, and linear trends are removed before correlation analysis.

## 4.2 Methods

### 4.2.1 Phase Locking Value

The Phase Locking Value (50) (PLV) is a nonlinear measure that quantifies how consistently the instantaneous phase of two oscillatory signals vary over time, independent of their amplitude or linear correlation.

To obtain the instantaneous phase of a real-valued signal $x(t)$ varying with time $t$, the signal is transformed into a complex-valued analytic signal $z(t)$ using the Hilbert transform:

$$z(t) = x(t) + i\mathcal{H}[x(t)], \tag{1}$$

where $\mathcal{H}[x(t)]$ denotes the Hilbert transform of $x(t)$.

The analytic signal can be expressed in polar form as

$$z(t) = A(t)e^{i\phi(t)}, \tag{2}$$

Where $A(t) = |z(t)| = \sqrt{(x(t))^2 + (\mathcal{H}[x(t)])^2}$ represents the instantaneous amplitude, and

$$\phi(t) = \arg\big(z(t)\big) = \tan^{-1}\left(\frac{\mathcal{H}[x(t)]}{x(t)}\right), \tag{3}$$

represents the instantaneous phase.

The PLV between two oscillatory signals $x_1(t)$ and $x_2(t)$ containing N data points is then computed as

$$\text{PLV} = \left|\frac{1}{N}\sum_{t=1}^{N} e^{i\Delta\phi(t)}\right|, \tag{4}$$

where $\Delta\phi(t) = \phi_1(t) - \phi_2(t)$, and $\phi_1(t)$ and $\phi_2(t)$ are the instantaneous phases of $x_1(t)$ and $x_2(t)$, respectively.

The PLV ranges from 0 to 1, where 0 indicates no phase synchrony and 1 indicates perfect phase locking between the two signals.

### 4.2.2 Phase space reconstruction

According to Takens' embedding theorem (33), the attractor of an unknown dynamical system can be reconstructed from time-delayed copies of a single observable, provided appropriate embedding parameters are chosen.

Given a scalar time series x(t), the reconstructed state vector at time t is defined as

$$X(t) = [x(t), x(t+\tau), x(t+2\tau), \ldots., x(t+(m-1)\tau)], \tag{5}$$

where $\tau$ is the time delay and m is the embedding dimension. The embedding dimension (m) and time delay (tau) are estimated individually for each time series by Cao's method (79) and average mutual information (80), respectively. Proper selection of (m, tau) ensures sufficient dimensionality to unfold the attractor without false projections.

Each X(t) represents a single point in an m-dimensional reconstructed phase space. As t evolves, a sequence of such vectors is generated, and the collection of all reconstructed vectors forms the trajectory of the system in the reconstructed phase space, approximating the underlying attractor.

The delay τ determines the temporal separation between coordinates and ensures that successive components contain new dynamical information, while the embedding dimension m specifies the number of delayed coordinates required to unfold the attractor without self-overlaps. When these parameters are appropriately chosen, the reconstructed phase space is topologically equivalent to the original system's attractor, enabling the investigation of nonlinear properties such as recurrence structure, determinism, and predictability.

#### 4.2.3 Joint Recurrence Plot

A recurrence plot provides a geometric 2-dimensional representation of how a dynamical system revisits similar states in its reconstructed phase space over time. Each white dot in the plot indicates that the system at time i is closer to its state at time j than a prescribed threshold distance (ε) in the phase space. The overall density of the white dots, quantified by the recurrence rate, represents the percentage of phase-space state pairs that recur within a given threshold distance (ε; see method). For example, if a system's evolution in phase space is represented by 10 phase space states, the recurrence plot compares all possible pairs of points, giving 10 x 10 = 100 comparisons. A recurrence rate of 20% means that 20 out of these 100 point pairs are close to each other in phase space i.e., their separation is smaller than the chosen threshold distance (ε), determined based on the recurrence rate.

A recurrence plot is constructed by identifying pairs of states that recur close to each other in phase space (81). The recurrence matrix is defined as

$$R_{ij}(\varepsilon) = \Theta\left(\varepsilon - \left\|X_i - X_j\right\|\right), \tag{6}$$

where $X_i \in \mathbb{R}^m, i = 1, 2, \ldots\ldots, N$ indicates sequence of N state vectors in phase space, $\|\cdot\|$ denotes the Euclidean norm, ε is the recurrence threshold, $\Theta(\cdot)$ is the Heaviside step function.

$$\Theta(u) = \begin{cases} 1, & u \geq 0, \\ 0, & u < 0. \end{cases} \tag{7}$$

The recurrence threshold (ε) and recurrence rate (RR) are related by the equation

$$RR(\varepsilon) = \frac{1}{N^2} \sum_{i,j=1}^{N} \Theta\left(\varepsilon - \|X_i - X_j\|\right) \tag{8}$$

Thus, for two dynamical systems X and Y, with recurrence matrices $R_{ij}^{x}$ and $R_{ij}^{y}$, the joint recurrence plot (JRP) is obtained as

$$JRP_{ij} = R_{ij}^{x} . R_{ij}^{y} \tag{9}$$

Hence, $JRP_{ij} = 1$ only when both systems recur simultaneously in their respective phase space.

The joint recurrence rate (JRR) quantifies the density of simultaneous recurrences in the joint recurrence plot (81). For two dynamical systems X and Y, JRR is obtained as

$$JRR(\varepsilon_x, \varepsilon_y) = \frac{1}{N^2} \sum_{i,j=1}^{N} \Theta\left(\varepsilon_x - \|X_i - X_j\|\right) . \Theta\left(\varepsilon_y - \|Y_i - Y_j\|\right) \tag{10}$$

**Acknowledgments:** DS and RIS acknowledge the funding from IndusInd Bank through the CSR project (CR23242519AEINIB002696) and the IOE initiative (Grant No. SP22231222CPETWOCTSHOC). BNG is grateful to Gauhati University, Guwahati for support through the ANRF Prime Minister Professor. GC acknowledges the seed grant IIT Delhi for the support.

**Funding:** This project was supported by IndusInd Bank through a CSR project (CR23242519AEINIB002696) and the Institute of Eminence (IoE) initiative of the Ministry of Human Resources and Development (MHRD) at IIT Madras; project number: SP22231222CPETWOCTSHOC.

**Data Availability:** Data related to this paper can be downloaded from:

COBE SST2 https://psl.noaa.gov/data/gridded/data.cobe2.html
HadISST https://climatedataguide.ucar.edu/climate-data/sst-data-hadisst-v11
ERSST https://psl.noaa.gov/data/gridded/data.noaa.ersst.v5.html
Kaplan SST https://psl.noaa.gov/data/gridded/data.kaplan_sst.html
IITM ISMR https://tropmet.res.in/static_pages.php?page_id=53
ERA5 ISMR https://cds.climate.copernicus.eu/datasets/reanalysis-era5-single-levels-monthly-means?tab=download
IMD ISMR https://www.imdpune.gov.in/lrfindex.php (Climate Monitoring > Gridded Data Archive > Rainfall (1.0 x 1.0) NetCDF)

IMD2 ISMR https://www.imdpune.gov.in/lrfindex.php (Climate Monitoring > SW Monsoon Rainfall Data)
SODA version 2.2.4 (for D20 and HC) http://apdrc.soest.hawaii.edu/datadoc/soda_2.2.4.php
CMIP6 https://esgf-node.llnl.gov/search/cmip6/
ERA5 meridional and vertical wind https://cds.climate.copernicus.eu/datasets/reanalysis-era5-pressure-levels-monthly-means?tab=download

**Author contributions:**
Conceptualization: DS, RIS, BNG
Methodology: DS, SS, RIS, BNG
Investigation: DS, RIS, BNG
Visualization: DS, SS, ST, GC, RIS, BNG
Supervision: RIS, BNG
Writing – original draft: DS, BNG
Writing – review & editing: DS, SS, ST, GC, RIS, BNG

**Competing interests:** Authors declare that they have no competing interests.

## Figures and Tables

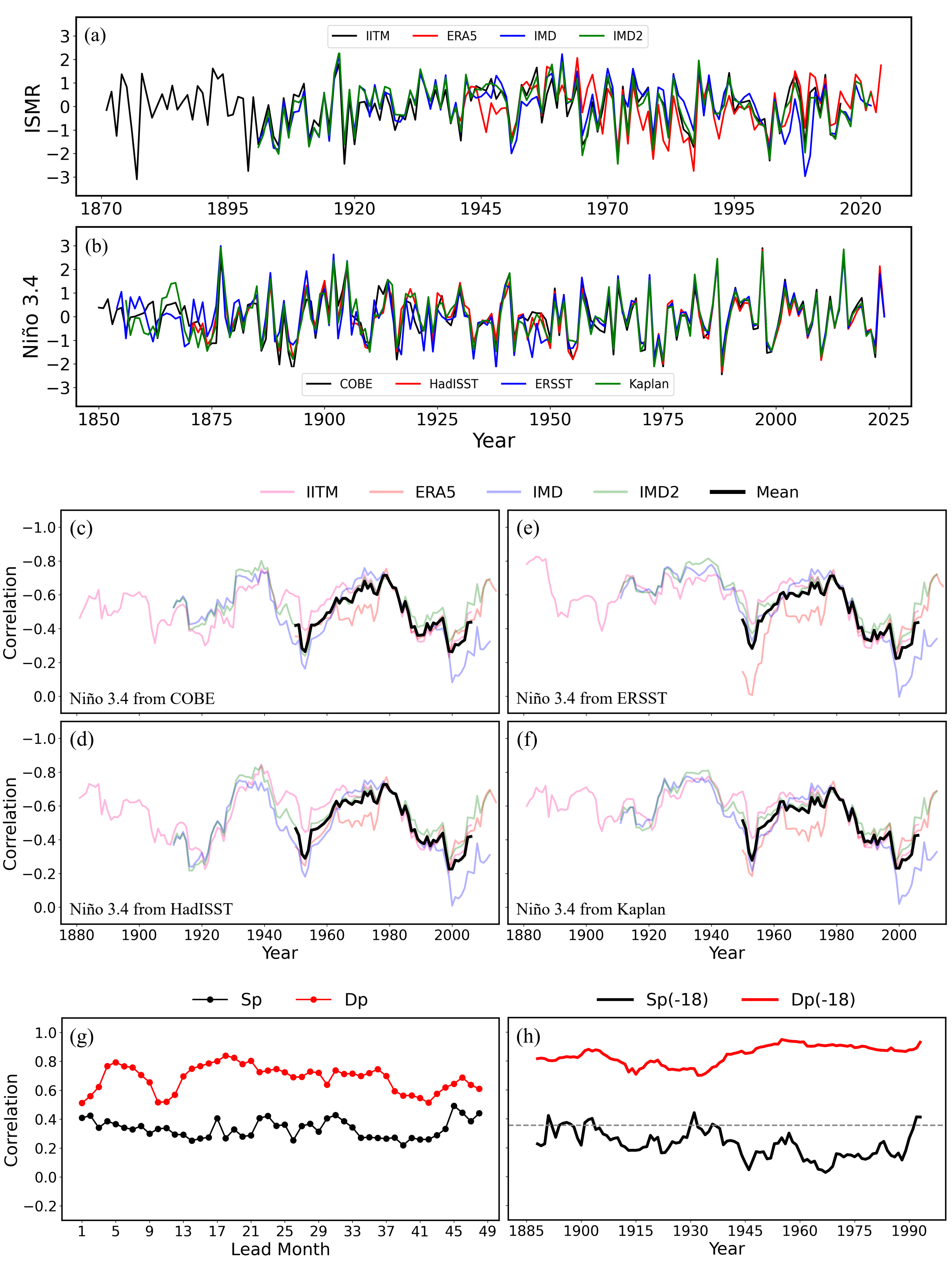

**Figure 1: Relation of ISMR with Pacific ENSO and Global ENSO. (a)** Time series of June-September mean normalized ISMR indices from four datasets (Indian Institute of Tropical Meteorology (IITM), the European Centre for Medium-Range Weather Forecasts Reanalysis version 5 (ERA5), India Meteorological Department (IMD), and IMD climate monitoring portal (hereafter referred as IMD2), **(b)** Time series of June-September mean normalized Niño 3.4 indices from four datasets Centennial In-situ Observation-Based Estimates of SSTs (COBE), Extended Reconstructed SST, Kaplan, and Hadley Centre Global Ice and SST (HadISST). 21-year moving simultaneous correlation between **(c)** June-September Niño 3.4 from COBE and ISMR index from (i) IITM, (ii) ERA5, (iii) IMD, and (iv) IMD2. The ensemble mean for the common period is shown by the thick line (black). **(d)** Same as (c) but with June-September Niño 3.4 from HadISST. **(e)** Same as (c) but with June-September Niño 3.4 from ERSST. **(f)** Same as (c) but with June-September Niño 3.4 from Kaplan. **(g)** Long-term (1875-2010) correlation of ISMR from IITM rainfall with Sp from COBE SST and Dp from Simple Ocean Data Assimilation (SODA) reanalysis at all leads up to 48-months. **(h)** 31-year moving correlation of ISMR with Dp and Sp at 18-month lead. The level of correlation for statistical significance at 95% confidence level is shown by the dashed line.

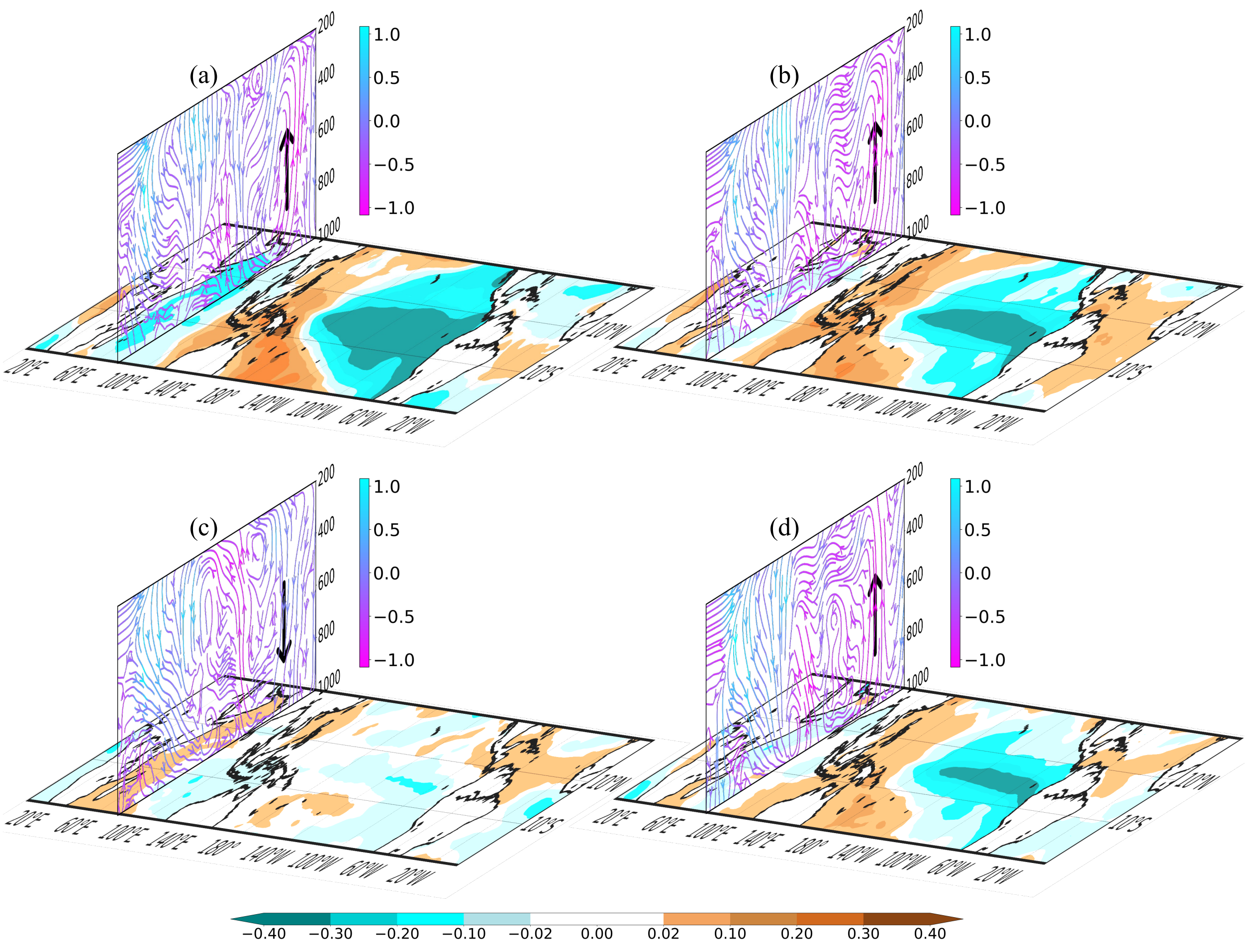

**Figure 2: Circulation pattern associated with the G-ENSO indices. (a)** Regression of Sp at lead 0 with June-September SST (shaded plot) and June-September meridional and vertical wind

anomalies averaged over 70°E-90°E from 1000-200 hPa (streamline plot). **(b)** Same as (a) but with Dp at lead 0, **(c)** Same as (a) but Sp at lead 18-months, **(d)** Same as (b) but Dp at lead 18-month. All time series are normalized by their respective grid wise standard deviation before the regression analysis. The colour of the streamline plots represent the regression of normalized Dp/Sp with normalized vertical wind anomalies, where negative/positive value indicates upward/downward motion (black arrow).

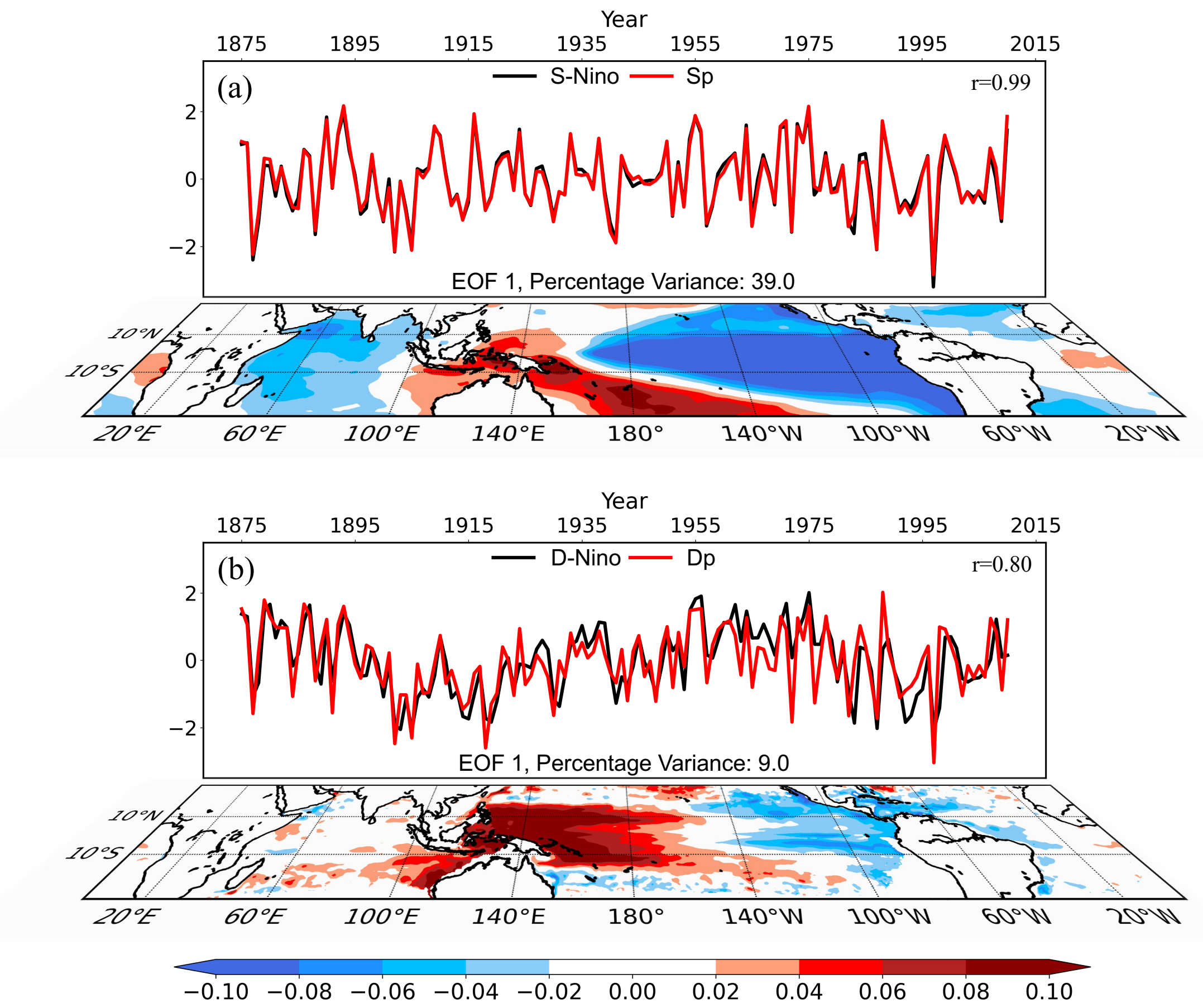

**Figure 3: Generalized G-ENSO indices. (a)** First empirical orthogonal function of June-September mean global tropical SST during 1875-2010 with percentage variance explained (bottom shaded plot). First principal component corresponding to the first empirical orthogonal function shown by the line plot at the top. First principal component of global tropical SST anomalies is referred to as S-Nino index. The correlation between S-Nino and ISMR-specific SST-based G-ENSO index (Sp) at lead zero is given by r. **(b)** Same as (c) but for June-September mean D20 anomalies. First principal component of global tropical D20 anomalies is referred to as D-Nino index. The correlation between D-Nino and ISMR-specific D20-based G-ENSO index (Dp) at lead zero is given by r.

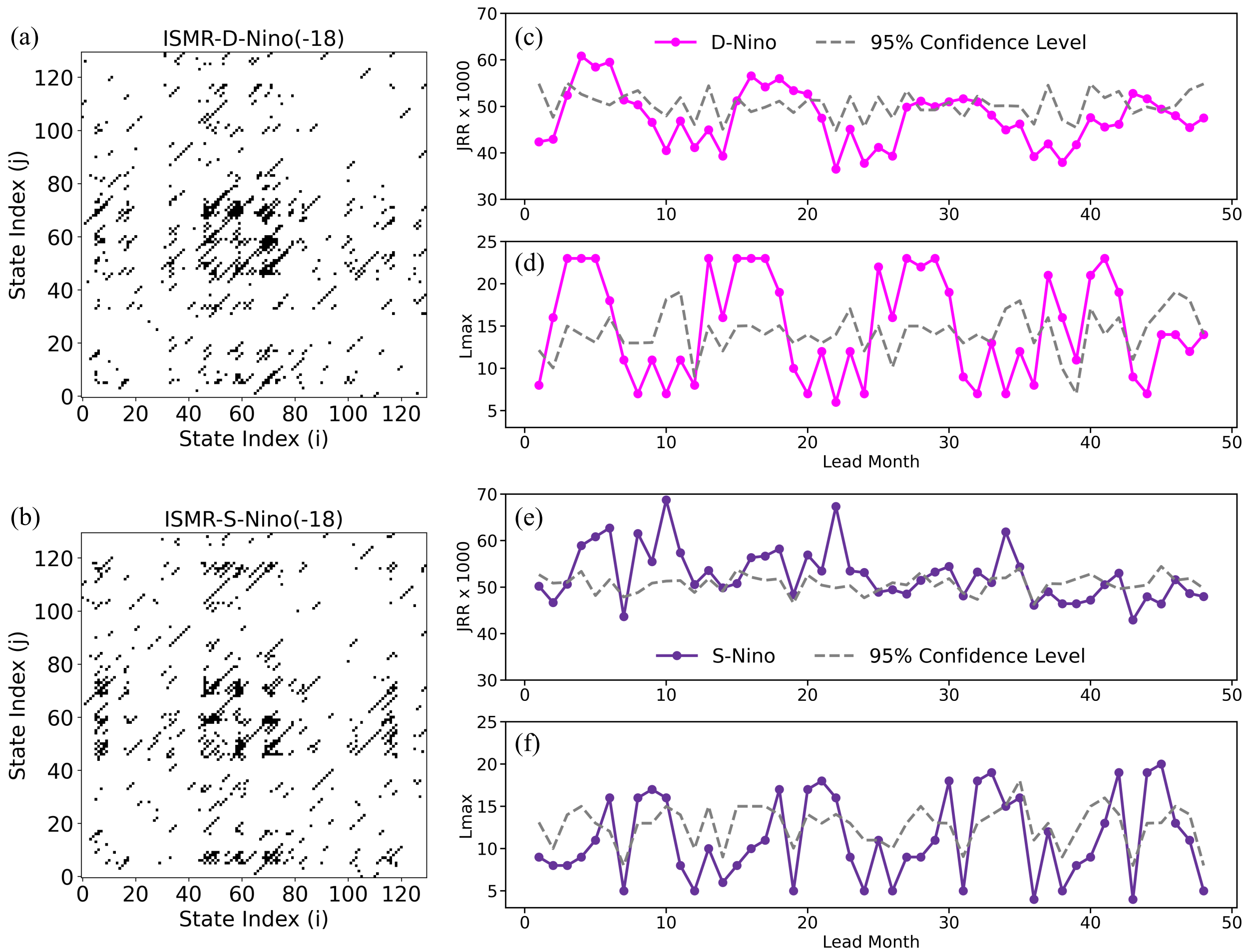

**Figure 4: Quantification of the coupling between ISMR and G-ENSO as a function of lead months. (a)** Joint recurrence plot constructed with 20% recurrence rate between the reconstructed phase space trajectory of ISMR and D-Nino at 18-month lead. **(b)** Same as (a) but with S-Nino at 18-month lead. **(c)** Quantification of joint recurrence rate between ISMR and D-Nino up to 48-month leads. **(d)** Length of maximum diagonal length in the joint recurrence plot between ISMR and D-Nino up to 48-month leads. **(e)** Same as (c) but for S-Nino. **(f)** Same as (d) but for S-Nino. Individual measures are scaled by appropriate powers of 10 for visual comparability. The dashed grey curves denote the values of the measures at 95% confidence level estimated from 100 iterative amplitude-adjusted Fourier transform (iAAFT) surrogate realizations. iAAFT preserves the amplitude distribution and power spectrum of the original time series while randomizing phase information (82, 83). This approach effectively destroys nonlinear temporal dependencies, allowing a robust test of whether the observed recurrence structures arise from true nonlinear deterministic dynamics rather than linear processes.

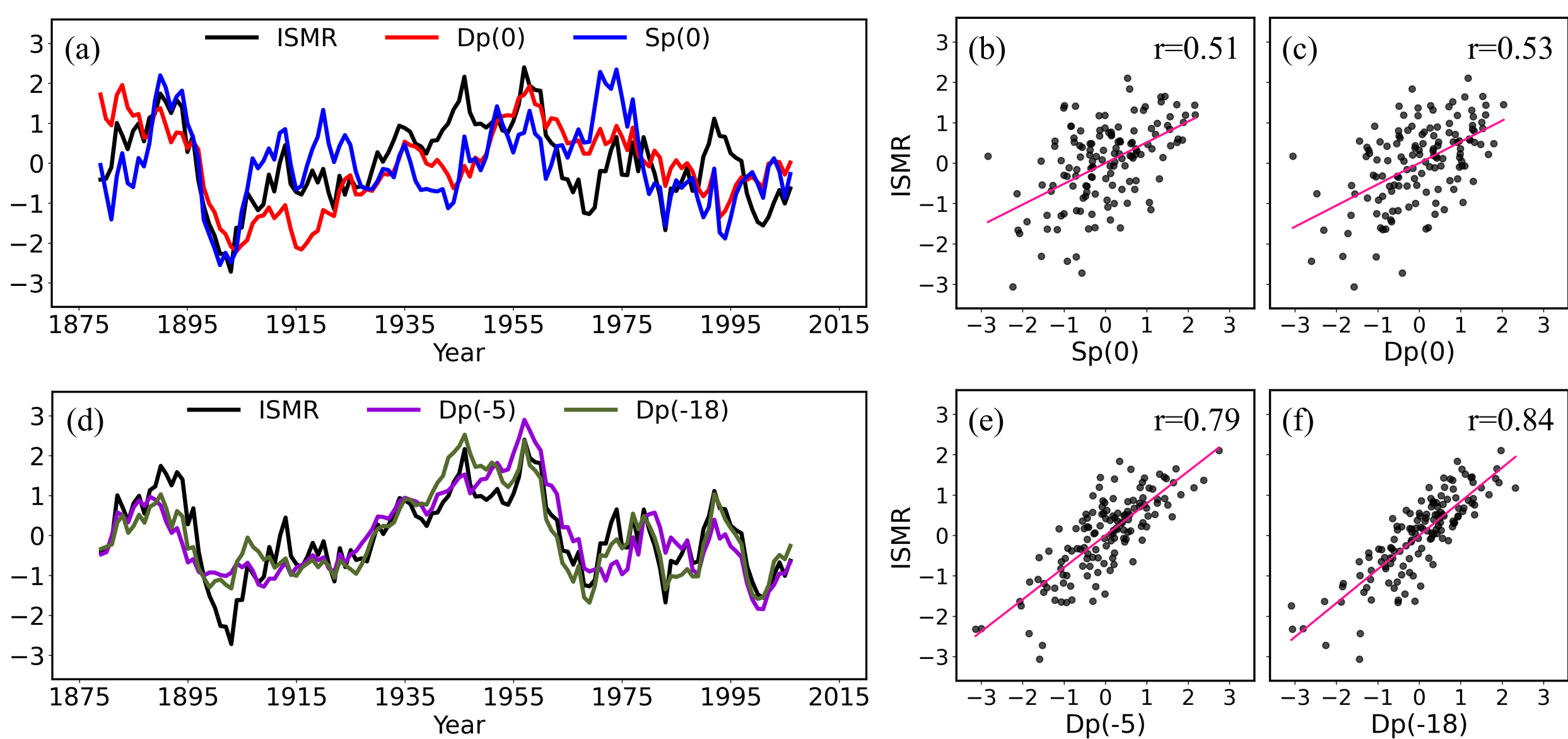

**Figure 5: Chaotic synchronization between ISMR and G-ENSO. (a)** 9-year moving average of ISMR from IITM data, Sp, and Dp at lead 0. **(b)** Scatter plot of ISMR with Sp(0). **(c)** Scatter plot of ISMR with Dp(0). **(d)** Same as (a) but for ISMR from IITM data and Dp at 5- and 18-month lead. **(e)** Scatter plot of ISMR with Dp(-5). **(f)** Scatter plot of ISMR with Dp(-18).

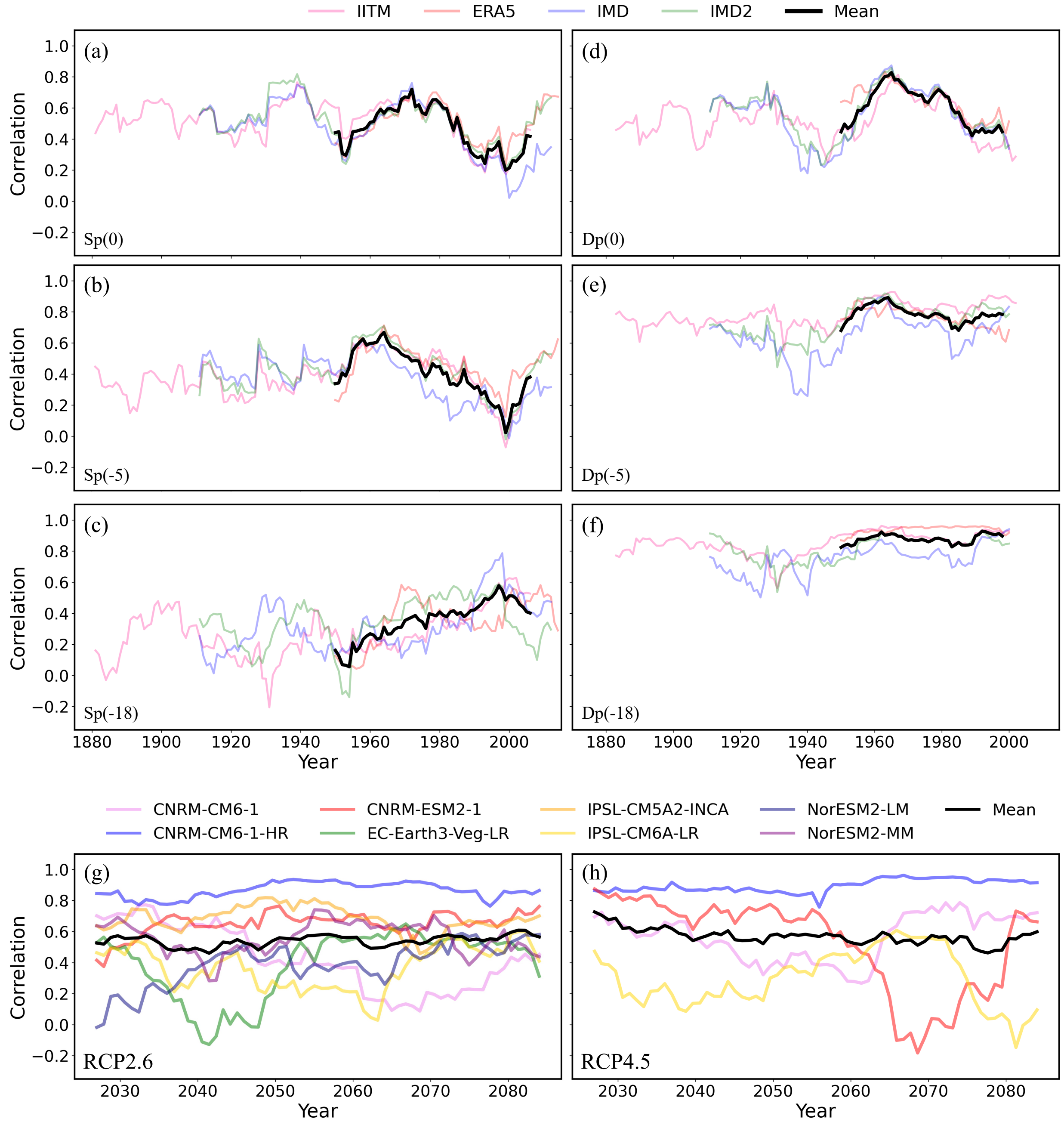

**Figure 6: G-ENSO-ISMR relationship in a warming climate. (a)** 21-year moving simultaneous correlation between June-September Sp and ISMR from (i) IITM, (ii) ERA5, (iii) IMD, and (iv) IMD2. **(b)** Same as (a) but with Sp at lead 5-month. **(c)** Same as (a) but with Sp at lead 18-month. **(d)** 21-year moving simultaneous correlation between June-September Dp and ISMR from (i) IITM, (ii) ERA5, (iii) IMD, and (iv) IMD2. **(e)** Same as (d) but with Dp at lead 5-month. **(f)** Same as (d) but with Dp at lead 18-month. The ensemble mean for the common period in (a-f) is shown by the thick line (black). SST and D20 datasets are obtained from COBE and SODA, respectively. 21-year moving correlation between Dp at 18-month lead and ISMR from CMIP6 model

simulations with **(g)** Representative Concentration Pathways 2.6 and **(h)** Representative Concentration Pathways 4.5. A detailed description of the CMIP6 models is shown in Table S2.

## Supporting Information for

## Resolving the Paradox of Changing El Niño-Monsoon Relation through Synchronization of Chaotic Oscillators

Devabrat Sharma[1,2,*], Sagar Sedani[1,2], Shruti Tandon[1,2], Gaurav Chopra[3], R. I. Sujith[1,2], and B. N. Goswami[4]

[1]Department of Aerospace Engineering, Indian Institute of Technology Madras, Chennai-600036, India
[2]Centre of Excellence for Studying Critical Transitions in Complex Systems, Indian Institute of Technology Madras, Chennai-600036, India
[3]Indian Institute of Technology Delhi, New Delhi-110016, India
[4]ST Radar Centre, Gauhati University, Guwahati-701014, India

*Devabrat Sharma
Email: devabratsharma9597@gmail.com

**This PDF file includes:**

**Text S1: ISMR-Specific G-ENSO Indices**

The Sp, and Dp indices are constructed to incorporate the simultaneous influence of all the potential tropical oceanic teleconnections associated with ISMR. The Sp index is derived from SST anomalies and the Dp index from D20 anomalies. All indices are computed over the tropical belt (0°–360°E, 30°S–30°N). For any given lead, the index is generated by projecting the anomaly field (SST for Sp and D20 for Dp) onto a long-term correlation pattern obtained between the corresponding anomaly field and ISMR anomalies during a reference period (Figure S1). Specifically, the correlation map is calculated using ISMR anomalies for the June–September (JJAS) season over a defined reference period (1875-2010, considered as lead 0). This correlation map represents the spatial pattern of oceanic variability most strongly linked to ISMR. To obtain the index at a particular lead, the SST (or D20) anomaly field every month at that lead is projected onto the statistically significant regions (>95% confidence level) of the correlation map derived from the reference period (Figure 1 of Sharma et al. (1)).

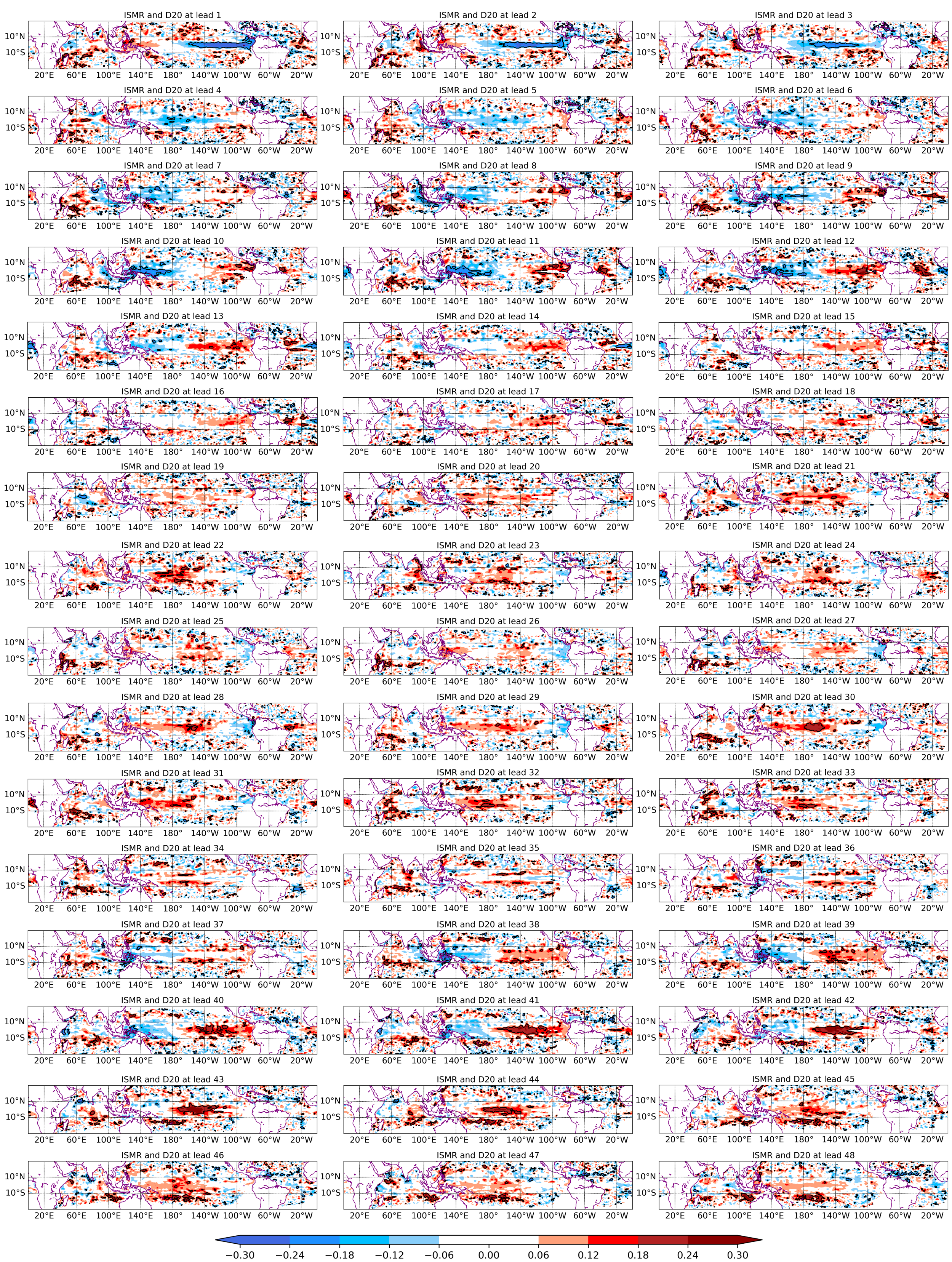

**Figure S1:** Correlation between ISMR and global tropical D20 anomaly from 1- to 48-month leads. The base period for over which ISMR is obtained is between 1875 and 2010. The dark contours represent correlation above 95% confidence level.

**Text S2: ISMR and G-ENSO in Phase Space**

Phase space reconstruction provides a framework to recover the underlying dynamics of multidimensional systems from a single time series (2, 3). The method is based on the principle that the evolution of a deterministic dynamical system is encoded in its temporal structure (3). Therefore, the multidimensional geometry of the system dynamics can be reconstructed by embedding the scalar time series into a higher dimensional space using time-delayed coordinates (3) (see Materials and Methods). Using appropriate spatial embedding and time delay, we reconstruct the phase space of ISMR, D-Nino, and S-Nino from their scalar time series (Figure S2).

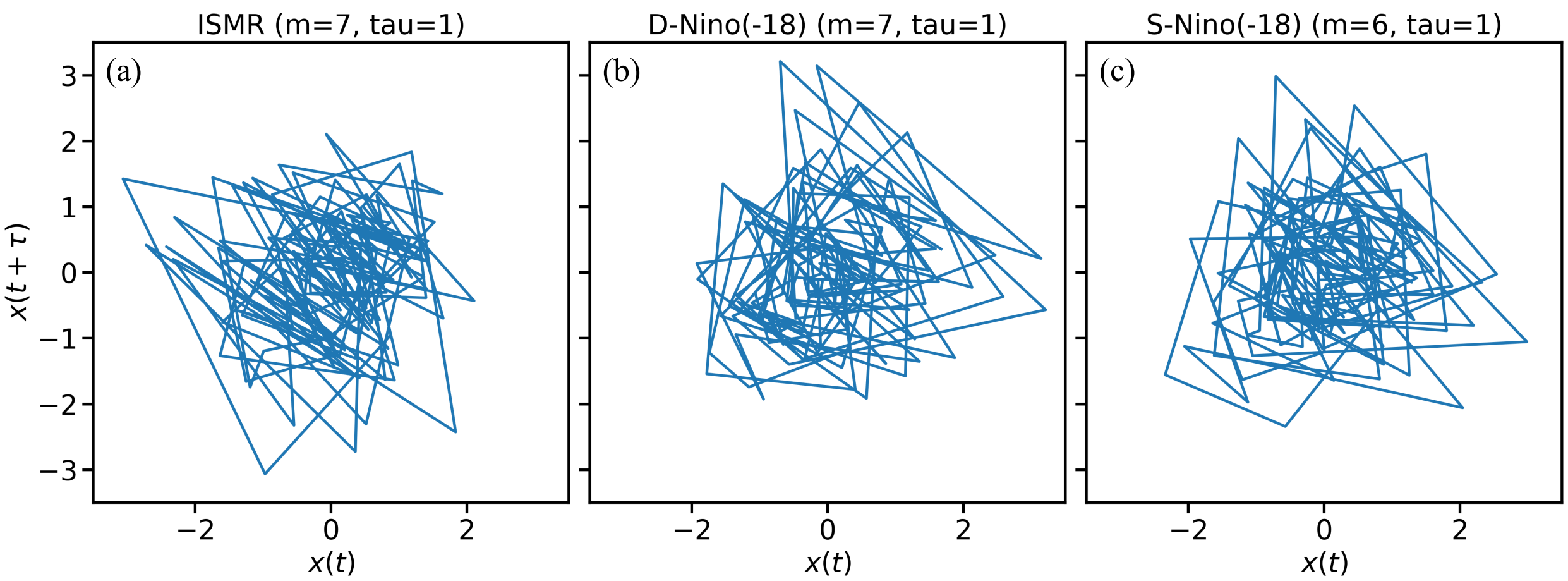

**Figure S2:** Phase space reconstruction of **(a)** ISMR time series between 1875 and 2010, **(b)** D-Nino time series at 18-month lead with respect to ISMR, and **(c)** S-Nino time series at 18-month lead with respect to ISMR.

The phase space of D-Nino and S-Nino are reconstructed for lead times up to 48-months relative to the monsoon season (Figure S2). Figure S2 indicates that the phase space trajectories of ISMR, D-Nino, and S-Nino are repeatedly stretched and folded within a bounded region of the phase space. Such geometry is characteristic of a strange attractor (4, 5), suggesting that the variability of these indices is governed by underlying nonlinear deterministic dynamics rather than purely stochastic processes. Individually, ISMR, D-Nino, and S-Nino behave as chaotic oscillators. As a result, the weak or insignificant correlation between D-Nino and ISMR, as well as S-Nino and ISMR, may arise because correlations quantify only pairwise linear associations and may therefore fail to capture nonlinear coupling and deterministic structure embedded in high dimensional climate systems.

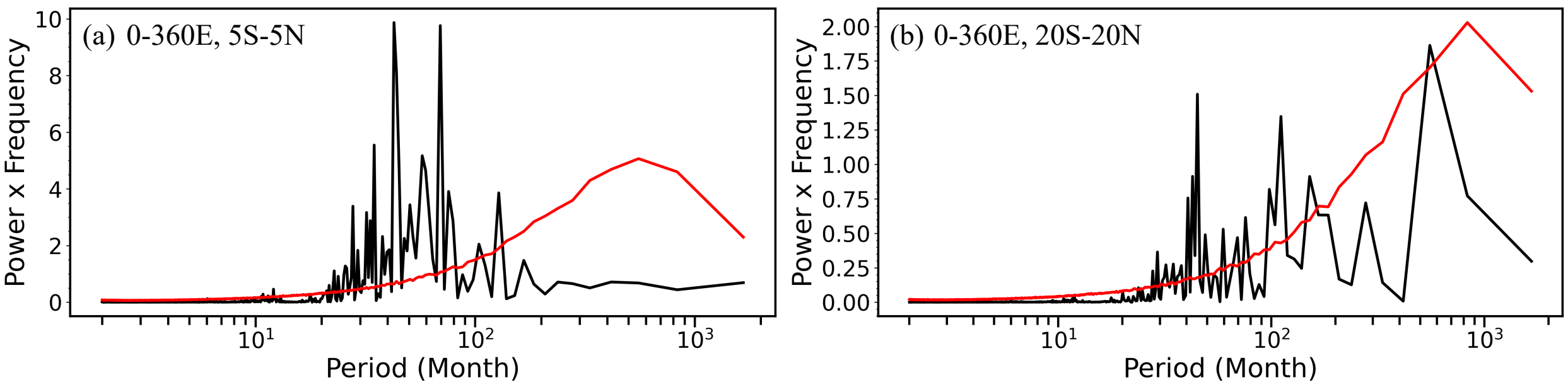

**Figure S3:** Power spectrum of 15-month running D20 anomaly averaged over the tropical region

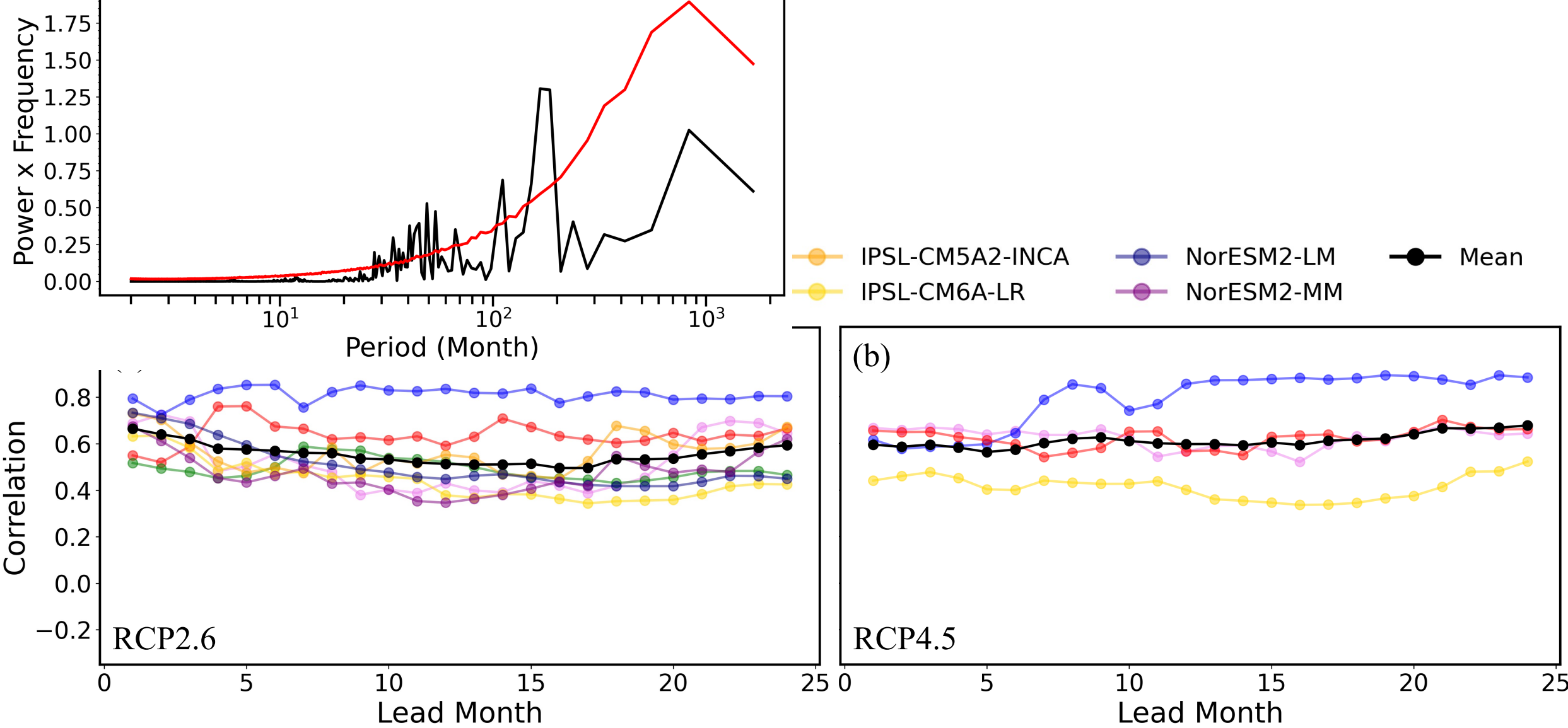

**Figure S4:** Correlation between Dp and ISMR from CMIP6 model simulations with **(a)** RCP2.6 and **(b)** RCP4.5. Lead zero period is over 2017-2100.

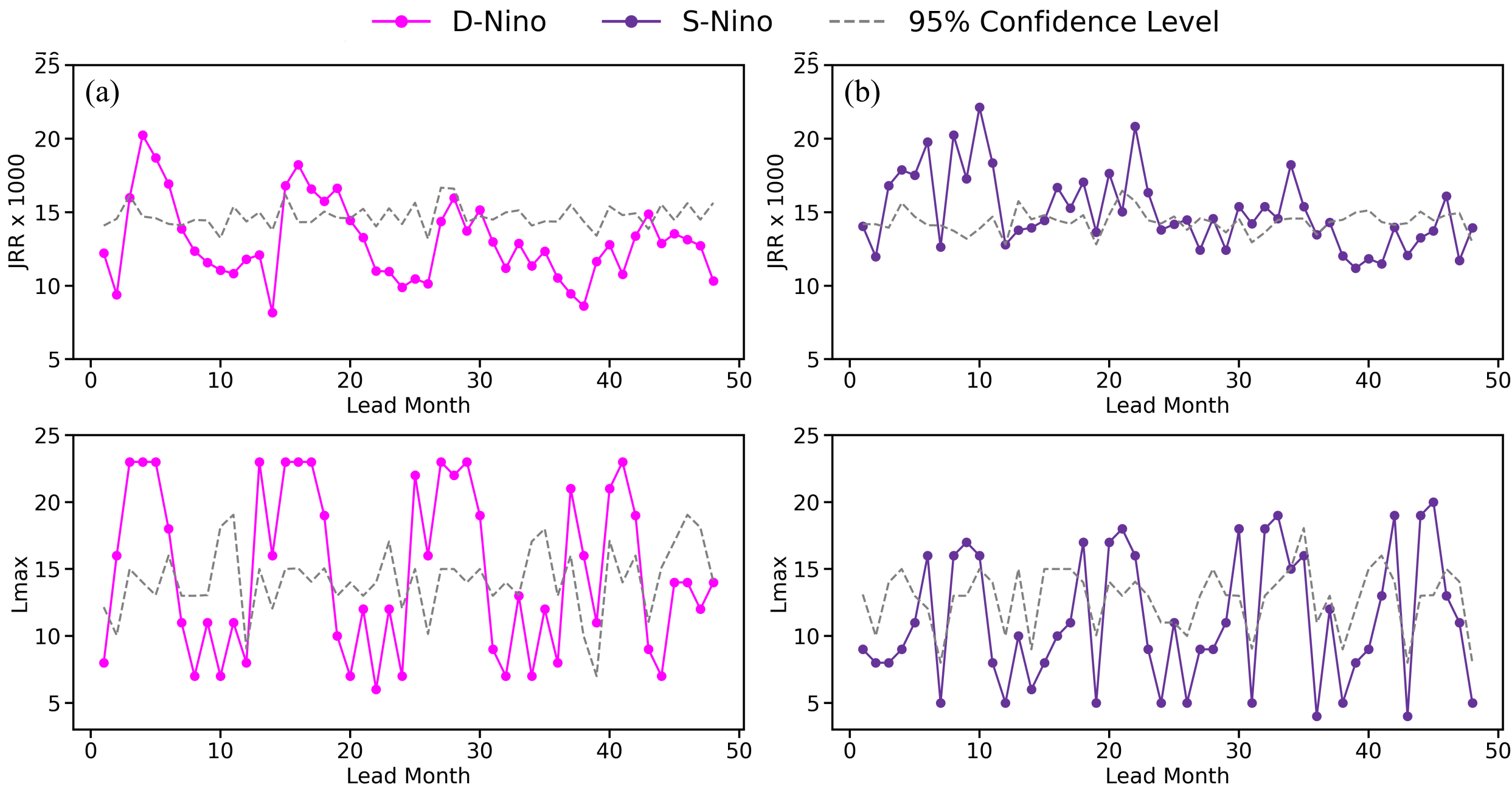

**Figure S5:** Quantification of joint recurrence plots (JRP) between ISMR and D-Nino and ISMR and S-Nino obtained at 10% recurrence rate. **(a-b)** joint recurrence rate (JRR), **(c-d)** maximum diagonal length (Lmax). Individual measures are scaled by appropriate powers of 10 for visual comparability. The dashed grey curves denote the values of the measures at 95% confidence level estimated from 100 iterative amplitude-adjusted Fourier transform (iAAFT) surrogate realizations.

**Table S1:** Details of the CMIP6 models used.

| Sl. No. | Model | RCP | Variable | Resolution (longitude x latitude) | Period |
|---|---|---|---|---|---|
| 1 | CNRM-CM6-1 | 2.6 and 4.5 | D20 and Rainfall | $1^{o}$ x $0.6^{o}$ | 2015-2100 |
| 2 | CNRM-CM6-1-HR | 2.6 and 4.5 | D20 and Rainfall | $0.25^{o}$ x $0.24^{o}$ | 2015-2100 |
| 3 | CNRM-ESM2-1 | 2.6 and 4.5 | D20 and Rainfall | $1^{o}$ x $0.6^{o}$ | 2015-2100 |
| 4 | EC-Earth3-Veg-LR | 2.6 | D20 and Rainfall | $1^{o}$ x $0.6^{o}$ | 2015-2100 |
| 5 | IPSL-CM5A2-INCA | 2.6 | D20 and Rainfall | $2^{o}$ x $1.28^{o}$ | 2015-2100 |
| 6 | IPSL-CM6A-LR | 2.6 and 4.5 | D20 and Rainfall | $1^{o}$ x $0.6^{o}$ | 2015-2100 |
| 7 | NorESM2-LM | 2.6 | D20 and Rainfall | $1^{o}$ x $0.38^{o}$ | 2015-2100 |
| 8 | NorESM2-MM | 2.6 | D20 and Rainfall | $1^{o}$ x $0.38^{o}$ | 2015-2100 |

**Table S2:** Mean and standard deviation of ISMR and Niño 3.4 from four datasets.

| Variable | Dataset | Period | Mean | Standard Deviation | Unit |
|---|---|---|---|---|---|
| **ISMR** | IITM | 1871-2016 | 848.16 | 83.17 | mm |
| | ERA5 | 1940-2024 | 915.07 | 89.01 | mm |
| | IMD | 1901-2022 | 942.43 | 93.48 | mm |
| | IMD2 | 1901-2022 | 887.76 | 86.80 | mm |
| **Niño3.4** | COBE | 1850-2024 | 26.87 | 0.63 | ºC |
| | ERSST | 1854-2024 | 26.68 | 0.72 | ºC |
| | HadISST | 1870-2024 | 26.99 | 0.61 | ºC |
| | Kaplan | 1856-2022 | - | 0.63 | ºC |